\documentclass[lettersize,journal]{IEEEtran}
\usepackage{amsmath,amsfonts}

\usepackage{array}
\usepackage[caption=false,font=normalsize,labelfont=sf,textfont=sf]{subfig}
\usepackage{textcomp}
\usepackage{stfloats}
\usepackage{url}
\usepackage{verbatim}
\usepackage{graphicx}
\usepackage{cite}
\usepackage{svg}
\usepackage[noend]{algpseudocode} 
\usepackage{etoolbox}

\usepackage{float}
\usepackage{pgfplots}
\usepackage[bookmarks,colorlinks]{hyperref}
\usepackage[linesnumbered,ruled,lined]{algorithm2e}
\usepackage{enumitem}
\usepackage{algpseudocode}
\usetikzlibrary{shapes.multipart,intersections}
\usepackage{cite}
\usepackage{amsmath,amssymb,amsfonts,amsthm,steinmetz}
\usepackage{graphicx}
\usepackage{mathrsfs}  
\usepackage{textcomp}
\usepackage{acronym}
\usepackage{xcolor}
\usepackage{upgreek,xspace}

\usepackage{tikz}
\usetikzlibrary{calc}
\makeatletter
\newcommand{\gettikzxy}[3]{%
  \tikz@scan@one@point\pgfutil@firstofone#1\relax
  \edef#2{\the\pgf@x}%
  \edef#3{\the\pgf@y}%
}
\makeatother

\usepackage[draft]{todonotes}

\usepackage{esvect}

\usepackage{times}
\usepackage{bm}
\usepackage{amsmath}
\usepackage{amssymb}
\usepackage{stmaryrd}
\usepackage{babel}
\usepackage{graphics, graphicx}
\usepackage{xcolor}
\usepackage{gensymb}
\usepackage{cite}
\usepackage{enumitem}
\usepackage{url}

\begin{document}

\title{Over-the-Air Emulation \\of Electronically Adjustable Rician MIMO Channels \\in a Programmable-Metasurface-Stirred \\Reverberation Chamber}

\author{Ismail~Ahmed,~Matthieu~Davy,~Hugo~Prod'homme,\\Philippe~Besnier,~\IEEEmembership{Senior Member,~IEEE},~and~Philipp~del~Hougne,~\IEEEmembership{Member,~IEEE}
\thanks{The authors acknowledge funding from the Rennes M\'{e}tropole acquisition d'\'{e}quipements scientifiques program (project ``SRI''), the European Union's European Regional Development Fund, and the French region of Brittany and Rennes Métropole through the contrats de plan État-Région program (project ``SOPHIE/STIC \& Ondes'').
}
\thanks{
The authors are with Univ Rennes, CNRS, CentraleSupélec, Nantes Université, IETR – UMR 6164, F-35000 Rennes.
}
\thanks{\textit{(Corresponding Author: Philipp del Hougne, philipp.del-hougne@univ-rennes.fr.)}}
}

\maketitle

\begin{abstract}
We experimentally investigate the feasibility of evaluating multiple-input multiple-output (MIMO) radio equipment under adjustable Rician fading channel conditions in a programmable-metasurface-stirred (PM-stirred) reverberation chamber (RC). Whereas within the ``smart radio environment’’ paradigm PMs offer partial control over the channels to the wireless system, in our use case the PM emulates the uncontrollable fading. We implement a desired Rician $K$-factor by sweeping a suitably sized subset of all meta-atoms through random configurations. We discover in our setup an upper bound on the accessible $K$-factors for which the statistics of the channel coefficient distributions closely follow the sought-after Rician distribution. We also discover a lower bound on the accessible $K$-factors in our setup: there are unstirred paths that never encounter the PM, and paths that encounter the PM are not fully stirred because the average of the meta-atoms’ accessible polarizability values is not zero (i.e., the meta-atoms have a non-zero ``structural'' cross-section). We corroborate these findings with experiments in an anechoic chamber, physics-compliant PhysFad simulations with Lorentzian vs “ideal” meta-atoms, and theoretical analysis. Our work clarifies the scope of applicability of PM-stirred RCs for MIMO Rician channel emulation, as well as electromagnetic compatibility test. 
\end{abstract}


\section{Introduction}

The trend of modern radio devices to use increasingly many integrated antennas inevitably requires a transition from conductive testing to over-the-air (OTA) testing in order to evaluate the radio equipment under realistic fading channel conditions~\cite{remley2017measurement,fan2018over,chen2018reverberation}. A fading condition of particular interest in signal processing and wireless communications is the Rician fading condition which assumes a complex Gaussian distribution of the channel coefficient fluctuations~\cite{parsons1992mobile,steele1999characterisation,vaughan2003channels}; the Rician $K$-factor quantifies the ratio of wave energy travelling via static vs dynamic paths. A well-established emulator of Rician fading conditions for the testing of wireless devices is the mechanically stirred reverberation chamber (RC)~\cite{holloway2006use,kildal2012characterization,remley2016configuring,chen2018reverberation}. An RC is an electrically large metallic cavity within which large metallic paddles are rotated to access different realizations of the dynamic paths~\cite{hill1998electromagnetic}. However, an electronic adjustment of the $K$-factor without requiring a cumbersome manual intervention and without altering any of the other channel characteristics is not possible with existing techniques, especially not for MIMO systems. 
For instance, replacing the paddles with differently sized ones would enable control over $K$ but implies a cumbersome and time-consuming manual intervention. Alternatively, by loading the RC with absorbing components (which usually also requires a manual intervention but has recently been proposed in an electronically controllable manner~\cite{yi2021switchable}), the $K$-factor can be controlled but its value is intertwined with the coherence bandwidth~\cite{chen2011estimation}. Finally, by using directive horn antennas and controlling their separation or relative orientation, the $K$-factor of a single-input single-output (SISO) link can be controlled quite easily, but the method does not generalize to MIMO systems. Similarly, adjusting the $K$-factor in post-processing by selecting a suitable subset of realizations from a well-stirred RC~\cite{de2023rician} does not generalize to MIMO systems.

In this paper, we experimentally investigate the extent to which stirring an RC with a programmable metasurface (PM) rather than a mechanical stirrer can enable the emulation of MIMO Rician fading conditions with electronically adjustable $K$-factor. The idea of electronically reconfiguring the boundary conditions of an RC to stirr the field inside the latter can be traced back to Ref.~\cite{klingler2005dispositif} (see also the discussion thereof in Sec.~2e of Ref.~\cite{serra2017reverberation}). Our present work constitutes the first full-scale investigation of this idea using a large PM (composed of $\Tilde{N}_\mathrm{S}=315$ 1-bit programmable meta-atoms, covering a surface of 1.44~$\mathrm{m}^2$) to electronically reconfigure the boundary conditions of a commercial RC (5.25~$\mathrm{m}^3$). By sweeping a subset of $N_\mathrm{S}\leq \Tilde{N}_\mathrm{S}$ programmable meta-atoms through random configurations, we access different realizations of the dynamic paths, and we control the $K$-factor by choosing the value of $N_\mathrm{S}$.
Previous related work explored electronic stirring via random PM configurations in contexts of wireless localization~\cite{del2018precise}, the tuning of a small metallic parallelepipedic box to wave chaos~\cite{gros2020tuning}, as well as Green's function retrieval~\cite{del2021diffuse}. However, all of these previous works were exclusively concerned with the stirred field component, which is a quantity obtained in post-processing from the measurable physical field by subtracting the latter's mean. While subtracting the mean physical field in post-processing can be reasonable in the aforementioned applications (assuming the measurements have a good dynamic range), applications in OTA channel emulation or electromagnetic compatibility (EMC) testing are concerned with the physical field. Moreover, subtracting the mean field precludes any investigation of electronically adjusting the $K$-factor.

While the $K$-factor is a simple and widespread metric to characterize Rician fading, it is easily overlooked that
\begin{enumerate}[label=(\roman*)]
    \item merely evaluating the $K$-factor does not guarantee that the underlying distribution is indeed Rician. A more rigorous analysis of the channel coefficients is required to confirm that their fluctuations follow a complex Gaussian distribution.
    \item the $K$-factor is itself a spatially distributed quantity. This fact is of particular importance in the present work concerned with emulating MIMO Rician fading conditions. While the choice of the value of $N_\mathrm{S}$ can control the mean of the distribution of the $K$-factors of the involved channels, the corresponding variance across the considered channels can be significant.
    \item neither the $K$-factor nor the Rician distribution are uniquely associated with a specific power-delay profile (PDP). The static paths are generally \textit{not} exclusively due to a direct line-of-sight (LOS) path; instead, significant unstirred multi-bounce paths contribute substantially to the static paths (see related discussion in Ref.~\cite{yeh2012first}). 
\end{enumerate}
In our present work, we rigorously analyze the distributions of channel coefficients [regarding (i)] and $K$-factors [regarding (ii)]. Since our analysis is focused on single-frequency channel coefficients, the issue from (iii) does not play a role in our analysis but specific applications of OTA channel emulation should check if they have requirements regarding the PDP.

Beyond OTA channel emulation, our work is also relevant to RC-based EMC tests where the use of PMs may be an enticing opportunity to replace mechanical stirrers, promising a larger working volume, faster switching between realizations, and potentially access to more independent realizations. While the small-scale experiments from Ref.~\cite{gros2020tuning} already suggested that the stirred field components in a large-scale PM-stirred RC may closely reach ideal statistical properties, it remains to be determined whether the static paths can be suppressed sufficiently. Indeed, EMC tests (particularly radiated immunity tests) are concerned with the physical fields (rather than a quantity extracted from them in post-processing), and require $K \rightarrow 0$, i.e., negligible static paths such that the Rician distribution specializes to the Rayleigh distribution. We identify two qualitatively distinct origins of unstirred wave energy in our work, thereby paving the way for future efforts aimed at eliminating them. On the one hand, wave energy travelling along paths that never encounter the PM remains unstirred; on the other hand, wave energy travelling along paths that encounter the PM is, in general, not fully stirred. The reverberation time plays a crucial role regarding the importance of both origins of unstirred wave energy.

Incidentally, the use of PMs (also known as reconfigurable reflectarrays, tunable impedance surfaces, spatial microwave modulators or reconfigurable intelligent surfaces, see Refs.~\cite{sievenpiper2003two,kamoda201160,kaina2014hybridized,cui2014coding}) is currently intensely investigated as part of a new ``smart radio environment'' paradigm~\cite{subrt2012intelligent,liaskos2018new,del2019optimally,renzo2019smart}. Therein, the PM endows wireless systems with some control over the wireless propagation channels, and significant algorithmic design efforts are currently invested into identifying optimized PM configurations for a desired wireless functionality. This use case of PMs qualitatively differs from the one we study in the present paper. We use the PM (or a subset of its meta-atoms) in random configurations (i.e.,  no algorithmic effort is required to optimize the PM configuration) in order to emulate the fading radio environment (i.e., the PM is \textit{not} under the wireless system's control).

The remainder of this paper is organized as follows.
In Sec.~\ref{sec_theory} we theoretically derive the two possible origins of static paths in PM-stirred RCs.
In Sec.~\ref{sec_Methods}, we describe our experimental setup and measurement procedure.
In Sec.~\ref{sec_Results}, we identify regimes with Rician statistics for our experimental setup, we estimate the accessible $K$-factor values therein, and we analyze the spatial distribution of the $K$-factor.
In Sec.~\ref{sec_additInvest}, we further investigate the origins of unstirred wave energy with an experiment in an anechoic chamber as well as various PhysFad~\cite{faqiri2022physfad} simulations.
We close in Sec.~\ref{sec_Conclusion} with a conclusion and an outlook to future work.

\textit{Notation.} The vector $\mathbf{a}$ containing the diagonal entries of the matrix $\mathbf{A}$ is denoted by $\mathbf{a} = \mathrm{diag}(\mathbf{A})$.

\section{Origins of Unstirred Wave Energy \\in PM-Stirred RCs}
\label{sec_theory}

\subsection{Generalities}
\label{subsec_gen}

We start with a theoretical analysis of the origins of unstirred wave energy in PM-stirred RCs based on a physics-compliant end-to-end model for PM-parametrized wireless channels under rich-scattering conditions that was recently proposed~\cite{faqiri2022physfad} and experimentally validated~\cite{sol2023experimentally} (see also the overview in the book chapter~\cite{BookChapter}). For the sake of completeness, we summarize its main features here and refer interested readers to Ref.~\cite{faqiri2022physfad} for additional background. The model, coined PhysFad, is derived from first physical principles and treats all wireless entities ($N_\mathrm{T}$ transmitting antennas, $N_\mathrm{R}$ receiving antennas, $\tilde{N}_\mathrm{S}$ programmable meta-atoms, $N_\mathrm{E}$ environmental scattering objects) as dipoles. The $i$th dipole is characterized by its polarizability $\alpha_i$, and the dipoles indexed $i$ and $j$ are coupled via the corresponding Green's function $G_{ij}$. The PM configuration determines the polarizabilities of the dipoles describing the programmable meta-atoms. An interaction matrix $\mathbf{W}\in\mathbb{C}^{N\times N}$ describes the system comprising $N = N_\mathrm{T}+N_\mathrm{R}+\tilde{N}_\mathrm{S}+N_\mathrm{E}$ dipoles. The diagonal and off-diagonal entries of $\mathbf{W}$ are, respectively, the inverse polarizabilities and negatives of the Green's functions:
\begin{equation}
W_{i,j}=\begin{cases}
\alpha_{i}^{-1}, & i=j\\
-G_{ij}, & i\neq j
\end{cases}.
\end{equation}
It is convenient to partition $\mathbf{W}$ into $4 \times 4$ blocks and then $2 \times 2$ blocks as follows:
\begin{equation}
\mathbf{W}=\begin{bmatrix}
\mathbf{W}_{\mathcal{TT}} & \mathbf{W}_{\mathcal{TR}} & \mathbf{W}_{\mathcal{TE}} & \mathbf{W}_{\mathcal{TS}}\\
\mathbf{W}_{\mathcal{RT}} & \mathbf{W}_{\mathcal{RR}} & \mathbf{W}_{\mathcal{RE}} & \mathbf{W}_{\mathcal{RS}}\\
\mathbf{W}_{\mathcal{ET}} & \mathbf{W}_{\mathcal{ER}} & \mathbf{W}_{\mathcal{EE}} & \mathbf{W}_{\mathcal{ES}}\\
\mathbf{W}_{\mathcal{ST}} & \mathbf{W}_{\mathcal{SR}} & \mathbf{W}_{\mathcal{SE}} & \mathbf{W}_{\mathcal{SS}}
\end{bmatrix} = \begin{bmatrix}
\mathbf{W}_{\mathcal{UU}} & \mathbf{W}_{\mathcal{US}}\\
\mathbf{W}_{\mathcal{SU}} & \mathbf{W}_{\mathcal{SS}}
\end{bmatrix},
\end{equation}
where $\mathcal{T}$, $\mathcal{R}$, $\mathcal{S}$ and $\mathcal{E}$ denote the sets of dipole indices corresponding to transmitting antennas, receiving antennas, programmable meta-atoms, and environmental scattering objects, respectively, and $\mathcal{U} = \mathcal{T} \cup \mathcal{R} \cup \mathcal{E}$. The PM configuration $\mathbf{c} \in \mathbb{C}^{\tilde{N}_\mathrm{S} \times 1}$  contains the inverse polarizabilities of the dipoles from set $\mathcal{S}$ and is hence encoded into the diagonal of the $\mathbf{W}_\mathcal{SS}$ block:
\begin{equation}
    \mathrm{diag}(\mathbf{W}_\mathcal{SS}) = \mathbf{c}.
\end{equation}

The end-to-end wireless channel matrix $\mathbf{H}\in \mathbb{C}^{N_\mathrm{R}\times N_\mathrm{T}}$ is proportional to the $\mathcal{RT}$ block of the inverse of $\mathbf{W}$~\cite{faqiri2022physfad}:
\begin{equation}
    \mathbf{H} \propto \left[ \mathbf{W}^{-1}\right]_\mathcal{RT}.
\end{equation}
It is the inversion of the interaction matrix that self-consistently captures the infinite number of increasingly long multi-bounce paths. It is instructive for our goal of understanding the origins of static paths to see this more clearly by expressing the matrix inversion as an infinite sum of matrix powers. Specifically, as detailed in Sec.~VII of Ref.~\cite{rabault2023tacit}, it follows from the block matrix inversion lemma that
\begin{subequations}
  \begin{equation}
     [\mathbf{W}^{-1}]_\mathcal{UU}  = \left( \mathbf{W}_\mathcal{UU} - \mathbf{W}_{\mathcal{US}} \mathbf{W}_{\mathcal{SS}}^{-1} \mathbf{W}_{\mathcal{SU}} \right)^{-1} 
\end{equation}
\begin{equation}
    =    \mathbf{W}_\mathcal{UU}^{-1} \sum_{k=0}^\infty \left( \mathbf{W}_{\mathcal{US}} \mathbf{W}_{\mathcal{SS}}^{-1} \mathbf{W}_{\mathcal{SU}} \mathbf{W}_\mathcal{UU}^{-1} \right)^k\label{eq5b}
\end{equation}
\begin{equation}
    =  \mathbf{W}_\mathcal{UU}^{-1} + \mathbf{W}_\mathcal{UU}^{-1}  \mathbf{W}_{\mathcal{US}} \mathbf{W}_{\mathcal{SS}}^{-1} \mathbf{W}_{\mathcal{SU}} \mathbf{W}_\mathcal{UU}^{-1} +\mathcal{O}( \mathbf{W}_\mathcal{SS}^{-2}),\label{eqrabC}
\end{equation}
\end{subequations}
where $\mathcal{O}( \mathbf{W}_\mathcal{SS}^{-2})$ denotes terms involving more than one matrix product with $\mathbf{W}_\mathcal{SS}^{-1}$, i.e., more than one interaction with the PM. As pointed out in Ref.~\cite{rabault2023tacit}, the common ratio $\mathbf{W}_{\mathcal{US}} \mathbf{W}_{\mathcal{SS}}^{-1} \mathbf{W}_{\mathcal{SU}} \mathbf{W}_\mathcal{UU}^{-1}$ of the infinite sum represents a bounce from $\mathcal{U}$ to $\mathcal{S}$ and back to $\mathcal{U}$.
The very first term in Eq.~(\ref{eqrabC}) (corresponding to $k=0$), $\mathbf{W}_\mathcal{UU}^{-1}$, captures all paths that never encounter the PM. This term itself can be decomposed as an infinite sum of increasingly long paths that bounce multiple times between the dipoles included in $\mathcal{U}$. The second term in Eq.~(\ref{eqrabC}), $\mathbf{W}_\mathcal{UU}^{-1}  \mathbf{W}_{\mathcal{US}} \mathbf{W}_{\mathcal{SS}}^{-1} \mathbf{W}_{\mathcal{SU}} \mathbf{W}_\mathcal{UU}^{-1}$, captures all paths that bounce once from $\mathcal{U}$ to $\mathcal{S}$ and back to $\mathcal{U}$. Recall that $\mathbf{W}_\mathcal{UU}^{-1}$ captures itself an infinite sum of increasingly long paths bouncing within $\mathcal{U}$. Similarly, the interaction with the PM itself, captured by $\mathbf{W}_{\mathcal{SS}}^{-1}$, can in turn be decomposed into an infinite sum of increasingly long paths bouncing between meta-atoms:
\begin{subequations}
\begin{equation}
      \mathbf{W}_{\mathcal{SS}}^{-1} = \left(\mathbf{\Phi}^{-1} +  \mathbf{\mathcal{M}}_{\mathcal{SS}}\right)^{-1} = \left( \mathbf{I}_{\tilde{N}_\mathrm{S}} +  \mathbf{\Phi}\mathbf{\mathcal{M}}_{\mathcal{SS}}\right)^{-1}   \mathbf{\Phi}
\end{equation}
\begin{equation}
    = \left(\sum_{k=0}^\infty 
\left( -\mathbf{\Phi}\mathbf{\mathcal{M}}_{\mathcal{SS}} \right)^{k}\right) \mathbf{\Phi}
\end{equation}
\begin{equation}
    = \mathbf{\Phi} -  \mathbf{\Phi}\mathcal{M}_\mathcal{SS} \mathbf{\Phi} + \mathcal{O}\left( \mathbf{\Phi}^{2} \right) 
\end{equation}
\end{subequations}
where $\mathbf{\Phi}^{-1} = \mathrm{diag}(\mathbf{c})$ and hence $\mathbf{\mathcal{M}}_{\mathcal{SS}} =  \mathbf{W}_{\mathcal{SS}} - \mathbf{\Phi}^{-1}$. Note that $\mathbf{\Phi}$ is a diagonal matrix whose entries are the meta-atom's polarizabilities. The analysis in terms of multi-bounce paths can be continued for the higher-order terms in Eq.~(\ref{eqrabC}) involving multiple bounces between $\mathcal{U}$ and $\mathcal{S}$.

\subsection{Two Qualitatively Different Origins of Unstirred Wave Energy}

For our goal of understanding the possible origins of unstirred wave energy, i.e., paths that are not affected by $\mathbf{c}$, a first clear contribution is apparent: the paths that never encounter the PM, corresponding to $k=0$ in Eq.~(\ref{eq5b}), are static by definition. To identify a further, maybe surprising, origin of unstirred wave energy, let us now consider a special scenario in which there are no paths that did not encounter the PM. Specifically, we assume that the only path from $\mathcal{T}$ to $\mathcal{R}$ is via $\mathcal{S}$ and all paths encounter the PM exactly once. This implies that the LOS path between $\mathcal{T}$ and $\mathcal{R}$ is blocked by an idealized perfect absorber, that the antenna dipoles minimially reflect incident waves, and that there are no environmental scattering objects (other than the idealized perfect absorber that is not treated as being composed of dipoles here). In that case, the expression for $\mathbf{H}$ simplifies to
\begin{equation}
    \mathbf{H} \propto \mathbf{W}_\mathcal{RS} \mathbf{W}_\mathcal{SS}^{-1} \mathbf{W}_\mathcal{ST} = \mathbf{W}_\mathcal{RS} \mathbf{\Phi} \mathbf{W}_\mathcal{ST} + \mathcal{O}(\mathbf{\Phi}^{2}).
\end{equation}
Now, we are interested in $\langle \mathbf{H} \rangle_\mathbf{c}$ to determine if there are any static paths that did encounter the PM. By inspection of the first term, $\mathbf{W}_\mathcal{RS} \mathbf{\Phi} \mathbf{W}_\mathcal{ST}$, it is apparent that $\langle \mathbf{H} \rangle_\mathbf{c} = \mathbf{0}$ is only possible if $\langle \mathbf{\Phi} \rangle_\mathbf{c} = \mathbf{0}$ which in turn is only possible if the average of the available polarizability values for the RIS is zero. Similarly, the higher order terms cannot average to zero unless the average of the available polarizability values for the RIS is zero. 

To summarize, there are in general two possible origins of unstirred wave energy in a PM-stirred RC:
\begin{enumerate}[label=(\roman*)]
\item Some waves travel along paths that never encounter the PM. The wave energy carried along those paths hence remains unstirred.
\item Upon an encounter with the PM, a portion of the wave energy remains unstirred if the available polarizability values of the meta-atoms do not average to zero.
\end{enumerate}
By interpreting the programmable meta-atoms as load-tunable backscatter antennas, the effect (ii) can also be understood as originating from the non-zero ``structural'' scattering cross-section of the meta-atoms defined for antennas in Refs.~\cite{king1949measurement,hansen1989relationships,hansen1990antenna}. 
The generality and implications of effect (ii) have so far not been widely recognized as contributing static paths to the PM-parametrized wireless channel matrix. 
Yet, most existing PM prototypes (including ours, as we evidence below) rely on relatively simple meta-atom designs with Lorentzian response whose phase and amplitude properties are intertwined such that the average of the accessible polarizability values is usually not zero, implying that the effect (ii) is significant. It is, however, possible to conceive more elaborate meta-atom designs that do offer (almost) independent phase and amplitude control (and thereby access to a set of polarizability values whose average is zero), such as the dual-resonance design recently prototyped in Ref.~\cite{sleasman2023dual}.

\textit{Remark:} The understanding developed in this section about the origins of unstirred paths is also relevant to wave control problems beyond PM-parameterized RCs, e.g., for threading light through dynamic complex media~\cite{mididoddi2023threading}.

\subsection{Relevant Factors to Suppress Static Paths}

Before closing Sec.~\ref{sec_theory}, we seek to get some insights into what parameters determine the importance of the two contributions to static paths. These insights pave the way for future efforts seeking to perfectly suppress all static paths to achieve Rayleigh fading. A pivotal role is played by the \textit{reverberation time}. The longer the waves dwell in the RC before being attenuated, the less important are the paths that did not encounter the PM relative to those that did. Moreover, more paths will encounter the PM more than once such that less unstirred wave energy is carried by paths that interacted with the PM. Thus, a higher dwell time will reduce both contributions to static paths identified above. With respect to the PM design, it is hence desirable to minimize absorption (to avoid reducing the reverberation time of the RC by inserting the PM), and to have access to polarizability values that average to zero. Moreover, having more programmable meta-atoms should generally reduce the portion of wave energy carried by static paths (unless the meta-atoms are strongly absorbing). Furthermore, using programmable meta-atoms with the largest possible scattering cross-section will help to reduce unstirred wave energy. Finally, we note that other factors like the relative positions of the various wireless entities (antennas, PM, environmental scattering objects) also play a role but this role is more difficult to quantify.

\section{Methods}
\label{sec_Methods}

\subsection{Experimental Setup}

\begin{figure}[t]
\centering
\includegraphics[width = \linewidth]{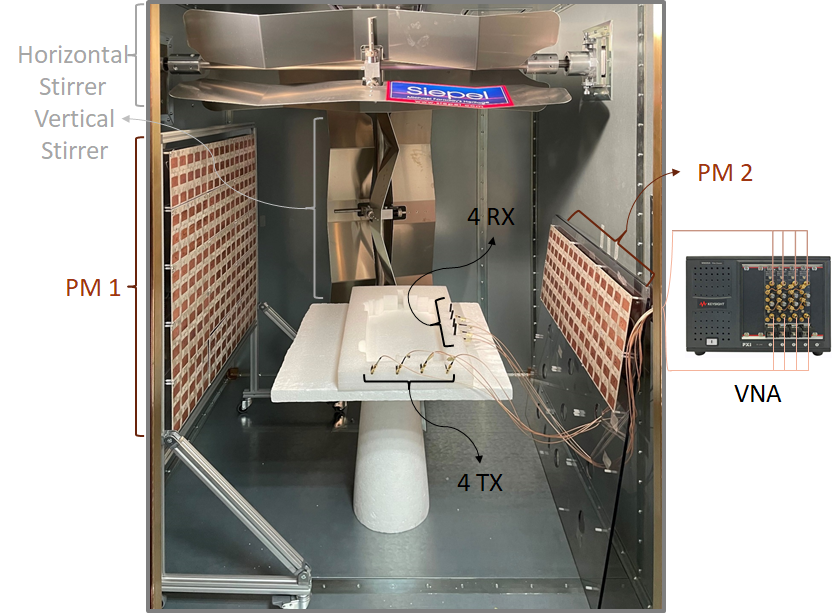}
\caption{Experimental setup involving a large-scale PM inside a commercial RC and a $4 \times 4$ MIMO setup based on commercial WiFi antennas. For our PM-based stirring experiments, the mechanical stirrers remain static throughout.}
\label{fig1}
\end{figure}

Our experiments are conducted within the commercial RC of dimensions $1.75\ \mathrm{m}\times 1.5\ \mathrm{m}\times × 2\ \mathrm{m}$ (volume: $5.25\ \mathrm{m}^3$; surface: $18.25\ \mathrm{m}^2$) shown in Fig.~\ref{fig1} that is equipped with a horizontal and a vertical mechanical stirrer (lowest usable frequency (LUF): 580~MHz). For our PM-based stirring experiments, the mechanical stirrers are not used and remain static throughout the experiment. 
Our PM prototype comprises 315 1-bit-programmable meta-atoms and has a total surface area of $1.44\ \mathrm{m}^2$. For practical purposes, these 315 meta-atoms are distributed across two mechanical supports, one with $15\times 15$ meta-atoms and one with $6\times 15$ meta-atoms. The meta-atom design follows that from Ref.~\cite{kaina2014hybridized} (except we use RO4003c laminate of 1.5~mm thickness, and the top microstrip is 6.5~mm rather than 9~mm long). These meta-atoms act on a single field polarization and seek to achieve in their two possible states a $\pi$ phase difference of the reflection coefficient under normal incidence at the central operating frequency of 2.445~GHz. The bandwith around this central frequency within which the meta-atoms significantly change the reflection coefficient is on the order of 100~MHz. The switching between the two possible states is achieved by controlling the bias voltage of a PIN diode embedded in the meta-atom. The interested reader is referred to Ref.~\cite{kaina2014hybridized} for additional details since the present work is concerned with the use of a PM for RC stirring rather than with designing a PM.

Our $4 \times 4$ MIMO system is built with eight commercial WiFi antennas (ANT-24G-HL90-SMA) designed for operation between 2.4~GHz and 2.5~GHz. As seen in Fig.~\ref{fig1}, the $N_\mathrm{T}=4$ ``transmitting'' antennas are aligned (i.e., in the same orientation) and regularly spaced by 6.5~cm (roughly half a wavelength); the $N_\mathrm{R}=4$ ``receiving'' antennas are similarly aligned and spaced, but their orientations are orthogonal to those of the ``transmitting'' antennas to avoid a direct LOS.
For a given realization of the PM configuration, we measure the $4 \times 4$ end-to-end channel matrix $\mathbf{H}$ using an eight-port vector network analyzer (VNA; Keysight~M9005A) for 3201 regularly spaced frequency points between 2~GHz and 3~GHz (intermediate-frequency bandwidth:~1~kHz; power:~10~dBm). The estimated signal-to-noise ratio of our measurements is 49.5~dB.

Based on the exponential decay of the channel impulse response envelope averaged over different antenna pairs and RC realizations (either using the mechanical stirrer or random PM configurations)~\cite{west2017best}, we estimate the composite quality factor of the RC in the vicinity of 2.5~GHz as $Q = 7639$ without the PM and $Q=1798$ with the PM. The drop in $Q$ suggests that the absorption of our PM prototype is significant compared to that of the RC alone. Based on Weyl's law, we estimate that the number of modes overlapping at any given frequency in the vicinity of 2.5~GHz is $n=8\pi V f_0^3 / c^3 Q \approx 42$ within the PM-equipped RC. 

\subsection{Measurement Procedure}
\label{subsec_measproced}

Our strategy to implement a desired $K$-factor is to sweep a subset of $N_\mathrm{S}$ meta-atoms through random configurations while the remaining $\tilde{N}_\mathrm{S}-N_\mathrm{S}$ meta-atoms are held in the ``OFF'' configuration throughout. To systematically investigate this approach, we consider the 13 different values of $N_\mathrm{S}$ summarized in the set
\begin{equation*}
    \mathcal{F} = \{1,10,20,32,63,95,126,157,188,220,251,282\}.
\end{equation*}
For each considered value of $N_\mathrm{S}$, we randomly select the $N_\mathrm{S}$ dynamic meta-atoms out of the $\Tilde{N}_\mathrm{S}$ available meta-atoms and measure $\mathbf{H}(f,\mathbf{c})$ for $m=1000$ random configurations of the selected $N_\mathrm{S}$ meta-atoms. We repeat this for five different random choices of the $N_\mathrm{S}$ meta-atoms for each value of $N_\mathrm{S}$.

\subsection{Estimation of $K$-factor}
\label{subsec_estimK}

Based on the acquired data, we estimate $K$ for each of the 16 channels for each choice of $N_\mathrm{S}$ for each of the five realizations. For the $(i,j)$th channel $H_{ij}$, the $K$-factor is theoretically defined as
\begin{equation}
    K(H_{ij}) = \frac{\mu_{ij}^2}{2\sigma_{ij}^2},
\end{equation}
where $\mu_{ij}$ and $\sigma_{ij}$ denote the absolute value of the mean and the standard deviation of the distribution of $H_{ij}$, respectively. In practice, we only have $m$ samples from the distribution of $H_{ij}$ in order to estimate its $K(H_{ij})$. Following Ref.~\cite{lemoine2010k} (see Eq.~(33) therein), the unbiased estimate $\hat{K}_{ij}$ of $K(H_{ij})$, assuming that the distribution is indeed Rician (see Sec.~\ref{subsec_estimGOF}), is given by
\begin{equation}
    \hat{K}_{ij} = \frac{m-2}{m-1} \frac{\hat{\mu}_{ij}^2}{2\hat{\sigma}_{ij}^2} - \frac{1}{m}, 
    \label{eq_philippe}
\end{equation}
where $\hat{\mu}_{ij}^2$ is the sum of the squares of the mean values of the real and imaginary parts of the available $m$ samples of $H_{ij}$, and $2\hat{\sigma}_{ij}^2$ is the sum of the variance of the real and imaginary parts of the available $m$ samples of $H_{ij}$. 
In the limit of $m\rightarrow \infty$, $\frac{m-2}{m-1} \rightarrow 1$ and $\frac{1}{m} \rightarrow 0$ such that $\hat{K}_{ij} \rightarrow \frac{\hat{\mu}_{ij}^2}{2\hat{\sigma}_{ij}^2}$ because the absolute value of the mean and the standard deviation of the samples approach the underlying distribution's true values thereof. An important implication of Eq.~(\ref{eq_philippe}) and Ref.~\cite{lemoine2010k} is that a meaningful estimation of $K$ in a scenario where its value is hypothesized to be very small requires a very large sample size $m$. For the range of values of $K$ that we report for our experiments in Sec.~\ref{sec_Results}, $m=1000$ is sufficient and the difference between $\hat{K}_{ij}$ and $\frac{\hat{\mu}_{ij}^2}{2\hat{\sigma}_{ij}^2}$ is negligible. However, for our simulations in Sec.~\ref{sec_additInvest}, we require significantly larger values of $m=5\times 10^6$. 

\subsection{Goodness-of-Fit Evaluation}
\label{subsec_estimGOF}

As pointed out in the introduction, a $K$-factor can be estimated for any set of measured samples, but a rigorous goodness-of-fit evaluation is necessary to determine whether the available $m$ samples are from a Rician distribution as opposed to from any other distribution. 
The choice of the utilized statistical test must be compatible with the fact that the distribution underlying our data is continuous, and that certain parameters of the distribution must be estimated from the available samples in our case. 
In sight of these considerations, we would like to use an Anderson-Darling (AD) test to evaluate the goodness of fit. The AD test is sensitive to extreme values of the distribution; it consists in evaluating a so-called $A^2$ metric and comparing it in terms of the modified metric $\tilde{A}^2 = A^2(1+\frac{0.6}{m})$ to a critical value $\mathcal{C}(\alpha)$, where $\alpha$ is the false-negative probability~\cite{lemoine2007investigation}.
However, a table of $\mathcal{C}(\alpha)$ has been worked out in Ref.~\cite{stephens1974edf} for the AD test applied to the Rayleigh distribution but not to the Rician distribution. 
Therefore, we perform an AD test regarding the hypothesis that the stirred field samples (i.e., the measured physical field samples minus their average) follow a Rayleigh distribution. To this end, we use the maximum likelihood method to estimate the parameters of the Rayleigh distribution from our $m$ samples after having subtracted the mean, in line with the assumption underlying the corresponding table of $\mathcal{C}(\alpha)$. The interested reader is referred to the Appendix of Ref.~\cite{lemoine2007investigation} for details regarding the implementation of this AD test. If the stirred field follows a Rayleigh distribution, then the physical field must follow a Rician distribution; if the stirred field does not follow a Rayleigh distribution, then the physical field cannot follow a Rician distribution. 

\section{Results}
\label{sec_Results}

\subsection{Identification of Regimes with Rician Statistics}

\begin{figure}[t]
\includegraphics[width=\linewidth]{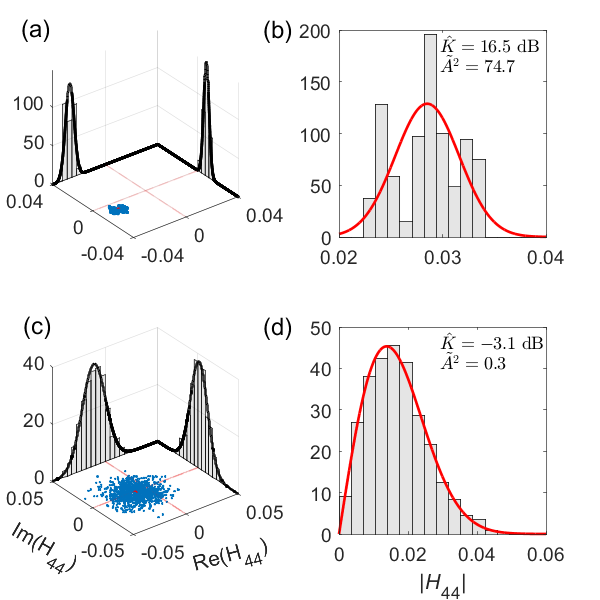}
\caption{For the channel coefficient $H_{44}$ at 2.445~GHz, we show in (a) [(c)] the fluctuations across $m=1000$ realizations for one choice of $N_\mathrm{S}=10$ [$N_\mathrm{S}=282$] dynamic meta-atoms (the other $\tilde{N}_\mathrm{S}-N_\mathrm{S}$ meta-atoms are held in their ``OFF'' state in all realizations). In addition, we show the probability density functions (PDFs) of $\mathrm{Re}(H_{44})$ and $\mathrm{Im}(H_{44})$ whose standard deviations are $3.2\times 10^{-3}$ and $2.7\times 10^{-3}$ [$1.1\times 10^{-3}$ and $1.1\times 10^{-2}$], respectively. The black line shows the best Gaussian fit. PDFs of $|H_{44}|$ are shown in (b) [(d)], and the corresponding AD metric $\Tilde{A}^2$ as well as the estimate $\hat{K}_{44}$ are indicated. The red line shows the best Rician fit.}
\label{fig2}
\end{figure}

To start, we analyze two representative sets of measured channel coefficient samples. Specifically, for the channel coefficient $H_{44}$ (i.e., linking the transmitter indexed 4 to the receiver indexed 4) at 2.445~GHz we analyze in Fig.~\ref{fig2}(a,b) and Fig.~\ref{fig2}(c,d) the distribution of $m=1000$ realizations of random configurations of one random choice of $N_\mathrm{S} = 10$ or $N_\mathrm{S} = 282$ dynamic meta-atoms, respectively. In the case of $N_\mathrm{S} = 10$, the distribution is not following Rician statistics. Upon visual inspection, we see that the cloud of $H_{44}$ values in Fig.~\ref{fig2}(a) is not isotropic and circular, in contrast to a complex Gaussian distribution. Moreover, the standard deviations of the real and imaginary parts of $H_{44}$ are significantly different ($3.2\times 10^{-3}$ vs $2.7\times 10^{-3}$). In terms of the distribution of the magnitude of $H_{44}$ shown in Fig.~\ref{fig2}(b), the AD-test fails, yielding a value of {$\tilde{A}^2=74.7>\mathcal{C}(0.05) = 1.341$}. Therefore, even though we can compute an estimate of $\hat{K}_{44} = 16.6\ \mathrm{dB}$, the distribution is not Rician in this case. In contrast, in the case of $N_\mathrm{S} = 282$, the samples are well described by a Rician distribution: the cloud in the complex plan is isotropic and circular upon visual inspection, real and imaginary parts have the same standard deviation of $1.1\times 10^{-2}$, and the AD-test is easily passed: {$\tilde{A}^2=0.3<\mathcal{C}(0.05) = 1.341$}. In this case, we estimate $\hat{K}_{44}=-3.1\ \mathrm{dB}$. The preliminary analysis from Fig.~\ref{fig2} highlights that the samples do not automatically always follow a Rician distribution, pointing to the need for a systematic check of the goodness of fit. 

\begin{figure}[t]
    \subfloat{\includegraphics[width=\linewidth]{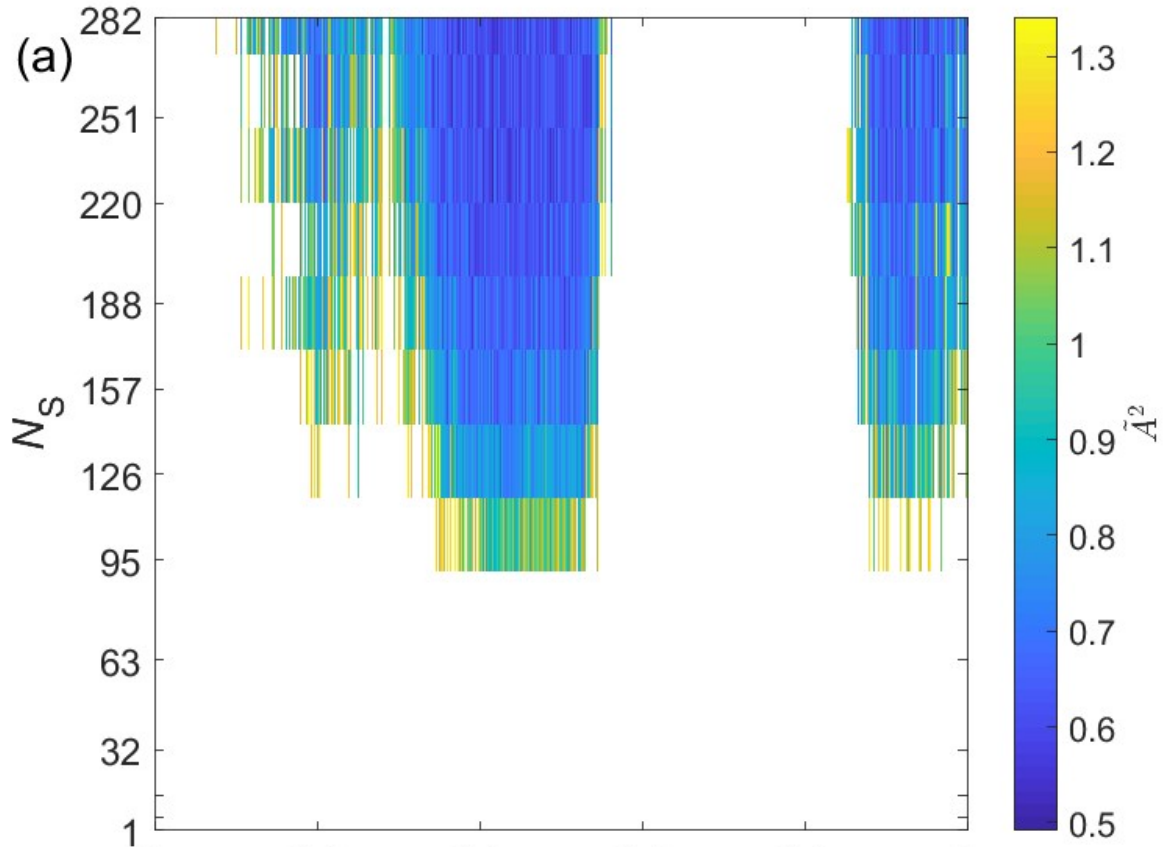}}
    \hspace{2pt}
    \subfloat{\includegraphics[width=\linewidth]{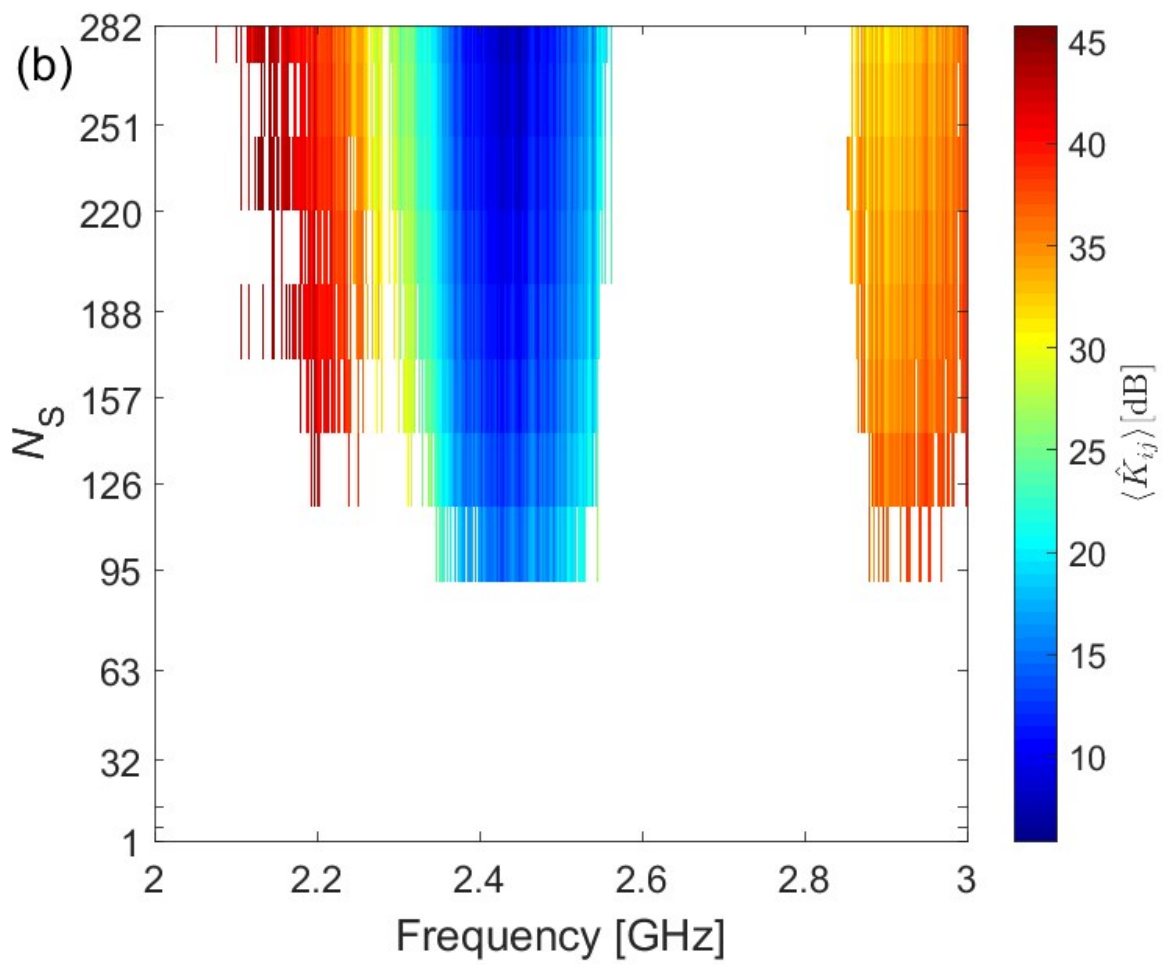}}
    \caption{(a) Colormap of the goodness-of-fit metric $\tilde{A}^2$ from the AD test as a function of frequency (horizontal axis) and $N_\mathrm{S}$ (vertical axis). The shown values are the average over the values obtained for the 16 channels and five different random choices of the $N_\mathrm{S}$ dynamic meta-atoms. The upper limit of the colorscale corresponds to $\mathcal{C}(0.5)=1.341$. (b) Corresponding values of $\hat{K}_{ij}$, again averaged over the 16 channels and 5 realizations.}
    \label{fig3}
\end{figure}

We now perform a systematic goodness-of-fit evaluation for all considered values of $N_\mathrm{S}$ and all considered frequency points. In Fig.~\ref{fig3}(a), we plot the value of $\tilde{A}^2$, averaged over the 16 channels of our $4 \times 4$ MIMO system and over five different choices of the $N_\mathrm{S}$ meta-atoms, as a function of frequency and of the chosen value of $N_\mathrm{S}$. At our PM's central working frequency of 2.445~GHz, the distribution is Rician for $N_\mathrm{S}\geq 95$. In line with the result from Fig.~\ref{fig2}, we hence conclude that sweeping through random configurations of a PM only yields Rician statistics if a sufficient number of meta-atoms is involved. A similar but distinct result was previously reported in Fig.~2 of Ref.~\cite{gros2020tuning} where, using a different metric, the goodness-of-fit of the stirred component of channel coefficients in a small metallic parallelepipedic box equipped with a PM was studied as a function of how many randomly chosen meta-atoms were configured to the ``ON'' state. (Note that this measurement procedure from Ref.~\cite{gros2020tuning} differs from ours outlined in Sec.~\ref{subsec_measproced}). 

We furthermore see in Fig.~\ref{fig3}(a) how the goodness of fit varies with frequency. We observe three frequency bands in which the AD test is passed: within a roughly 200~MHz wide interval around the central working frequency if $N_\mathrm{S}\geq 95$, as well as within a roughly 100~MHz wide interval around 2.9~GHz if $N_\mathrm{S}\geq 126$ and for some frequencies around 2.2~GHz if $N_\mathrm{S}>158$. The two sidebands arise from the working principle of our meta-atoms based on hybridized resonances, as outlined in Ref.~\cite{kaina2014hybridized}, yielding either one resonance at the working frequency or two resonances on either side of the working frequency. The sideband frequencies cannot be straightforwardly predicted based on full-wave simulations of a single meta-atom under normal incidence because these do not take the coupling for a heterogeneous PM configuration into account, nor the diversity of the angles of incidence in an RC. The effect of the PM in the sidebands is very weak, and in addition our antennas are not efficiently radiating at the sideband frequencies, but since our measurements are performed with a high dynamic range, the field fluctuations due to PM-based stirring are captured nonetheless. 
Incidentally, the effect of the sidebands also manifested itself for a very different metric (wireless localization precision) and a slightly different PM prototype in Fig.~4(c) of Ref.~\cite{del2018precise}.

The fact that Rician statistics are obtained with large values of $N_\mathrm{S}$ in the sidebands but not with low values of $N_\mathrm{S}$ in the main band highlights that the pivotal factor with respect to achieving Rician statistics is not the strength of the perturbation but the fact that it has many spatially distributed origins. In other words, the rank of the change of the interaction matrix between different realizations should be large (see also related discussions about the change of the interaction matrix upon updating the PM configuration in the Appendix and Ref.~\cite{prod2023efficient}).

\subsection{Identification of Accessible Average $K$-Factors}
\label{subsec_access_av_K}

Having defined the pairs of frequency and value of $N_\mathrm{S}$ for which the measured samples are well described by a Rician distribution, we now compute the corresponding estimates of the $K$-factor, averaged again over the 16 channel coefficients and the five random choices of the $N_\mathrm{S}$ dynamic meta-atoms. The dependence of $\langle \hat{K}_{ij} \rangle$ on the frequency and the value of $N_\mathrm{S}$ is shown in Fig.~\ref{fig3}(b) for cases in which the distribution is Rician according to the AD test from Fig.~\ref{fig3}(a). Although the overall range of values of $\langle \hat{K}_{ij} \rangle$ seen in Fig.~\ref{fig3}(b) ranges from around 5~dB to 45~dB, at any given frequency the range of accessible values is much more restricted. Within the central band, the lowest values are achieved (since the PM has the largest impact on the field) but high values of $\langle \hat{K}_{ij} \rangle$ remain out of reach.
Meanwhile, in the sidebands, a much higher set of values of $\langle \hat{K}_{ij} \rangle$ is accessible (while satisfying Rician statistics).

Although the range of accessible values of $\langle \hat{K}_{ij} \rangle$ within the central band is comparable to that expected in realistic indoor scenarios \cite{park2010ricean,medawar2013approximate
,real_world_K}, overall the limited range is rather surprising in sight of recent enthusiasm about PM-stirred RCs for EMC tests. In addition, it is striking that $\langle \hat{K}_{ij} \rangle$ never reaches 0~dB or lower, excluding the possibility of getting close to Rayleigh fading conditions in our current setup, or using the latter for EMC tests. 

Since our present work is the first to analyze \textit{physical} fields in PM-stirred RCs (as opposed to the analysis of \textit{stirred} fields in PM-stirred RCs in prior works~\cite{del2018precise,gros2020tuning,del2021diffuse}), Fig.~\ref{fig3}(b) is a key contribution of this paper. Unfortunately, it does not allow for enthusiastic conclusions regarding the \textit{easy} implementation of arbitrary Rician fading for OTA tests in a PM-stirred RC nor the \textit{easy} suitability of PM-stirred RCs for EMC tests. Instead, it suggests that PM-stirred RCs are likely suitable for specialized OTA applications requiring Rician statistics with a significant portion of unstirred energy, and that it is unlikely that PM-stirred RCs can \textit{easily} be suitable for EMC tests. Although these findings baffle rather than fuel the hope for a wide potential application scope of PM-stirred RCs in OTA and EMC tests, clarifying these points is important for the OTA and EMC communities. Moreover, while it might not be \textit{easy} to approximate Rayleigh fading conditions, it may nonetheless be possible in future work, 
motivating our additional investigations about the origins of unstirred wave energy in Sec.~\ref{sec_additInvest}.

\subsection{Spatial Distribution of $K$-Factors}
 
\begin{figure}[t]
\includegraphics[width=1\linewidth]{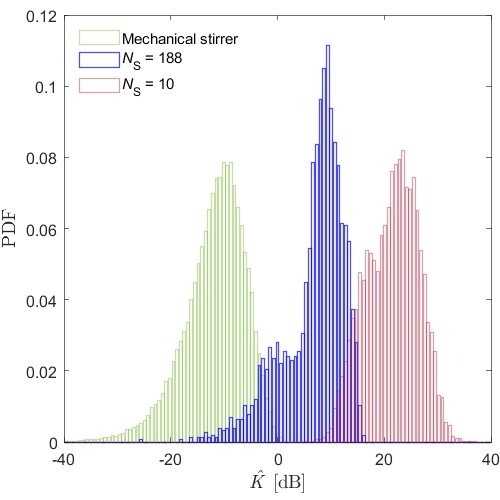}
\caption{PDFs of the estimated Rician $K$-factor $\hat{K}_{ij}$ at 2.445~GHz for three different stirring conditions: conventional mechanical stirring (green), PM-based stirring with $N_\mathrm{S}=188$ (blue) and PM-based stirring with $N_\mathrm{S}=10$ (red). In each case, the PDF is based on the values of $\hat{K}_{ij}$ for the 16 channels and 400 random choices of the $N_\mathrm{S}$ dynamic meta-atoms.}
\label{fig4}
\end{figure}

So far, we only looked at the mean of the $K$-factor estimates in Fig.~\ref{fig3}(b). However, $K$ is itself a spatially distributed quantity. The spread of $K$-factor values across the 16 considered channel coefficients is far from negligible, easily spanning one to two orders of magnitude, as shown in Fig.~\ref{fig4} for two representative values of $N_\mathrm{S}$. However, this spread is not specific to PM-based stirring but is also observed with mechanical stirring, as seen in Fig.~\ref{fig4}. Even though $\langle \hat{K}_{ij} \rangle$ never even reached 0~dB in Fig.~\ref{fig3}(b), for some realizations we observe values of $\hat{K}_{ij}=-20\ \mathrm{dB}$ and lower even with $N_\mathrm{S}=188$, as seen in Fig.~\ref{fig4}. Measurements in real-life scenarios reported in the literature similarly revealed large fluctuations of the estimated $K$-factors~\cite{real_world_K}.
Many RC based studies to date were only concerned with SISO setups such that they did not observe this spread, but our results in Fig.~\ref{fig4} raise doubts about the suitability of the $K$-factor to characterize the channel fluctuations beyond SISO settings.

\section{Additional Investigations \\Regarding Origins of Unstirred Wave Energy}
\label{sec_additInvest}

Our identification of the regimes with Rician statistics as well as the therein accessible $K$-factors for our experimental setup from Fig.~\ref{fig1} has revealed that significant amounts of unstirred wave energy impose a lower bound on the $K$-factors that we can access in our experimental setup. Both for OTA channel emulation as well as EMC testing, it is important to understand to what extent and how future work could significantly reduce the portion of wave energy not affected by the PM configuration. For instance, would it help to use more meta-atoms? To reduce the meta-atoms' absorption? To seek programmable meta-atoms whose accessible polarizability values average to zero? 

As developed theoretically in Sec.~\ref{sec_theory}, two qualitatively different contributions to unstirred wave energy exist in a PM-stirred RC: (i) wave energy travelling along paths that never encounter the PM, and (ii) portions of wave energy that remain unaffected by the PM configuration despite encountering the PM (due to the non-zero average of the accessible polarizability values). In this section, first, we experimentally evidence the importance of effect (ii) in an anechoic-chamber setup in which each path connecting the transmitter to the receiver has encountered the PM exactly once. Then, second, we further evidence the importance of effect (ii) numerically based on the open-source PhysFad simulator~\cite{faqiri2022physfad} by considering a setting analogous to the anechoic-chamber experiment but with idealized meta-atoms whose accessible polarizability values average to zero. Then, third, we evidence that reverberation can mitigate effect (ii) by numerically considering in PhysFad an RC setting in which the PM covers all walls of an RC. Finally, fourth, we explore in the PhysFad setting of a fully PM-cladded RC how the extent to which the meta-atoms absorb wave energy (and thereby reduce the reverberation) impacts effect (ii).

\subsection{Experimental Isolation of the Second Origin of Unstirred Wave Energy}
\label{subsec_AC}

\begin{figure}[h]
\includegraphics[width=\linewidth]{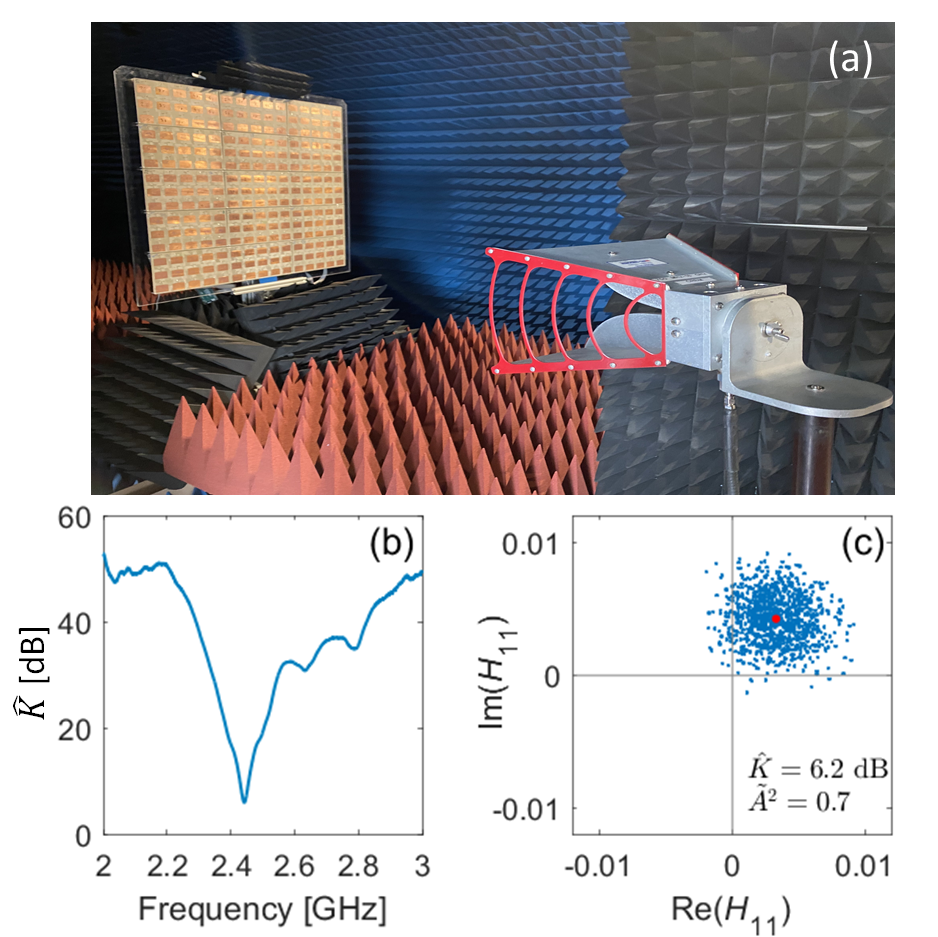}
\caption{PM-based stirring in an anechoic chamber (AC). (a) Experimental setup involving a large-scale PM inside an AC and two horn antennas (ETS-3115). The distance between the antennas and the PM is 2.3~m. (b) Estimated $K$-factor as a function of frequency. (c) Spread of channel coefficients $H_{11}$ in the complex plane. The corresponding AD metric $\Tilde{A}^2$ as well as the estimate $\hat{K}_{11}$ are indicated. }
\label{fig5}
\end{figure}

We now check experimentally for our PM prototype whether the second origin of unstirred wave energy in PM-stirred RCs identified in Sec.~\ref{sec_theory} (the fact that the accessible polarizability values do not average to zero) is significant by isolating it. To that end, we perform the experiment imagined in Sec.~\ref{sec_theory}: we place our PM in an anechoic chamber (AC) and we place absorbers between the two antennas such that the only significant path connecting $\mathcal{T}$ to $\mathcal{R}$ is via the PM -- see Fig.~\ref{fig5}(a). In other words, no significant fraction of the wave energy can avoid interacting with the PM. Moreover, since the antennas are not creating strong reflections, no significant paths encounter the PM more than once. In other words, all paths from transmitter to receiver encounter the PM exactly once.
If the second effect identified in Sec.~\ref{sec_theory} was negligible, we should now observe $K \rightarrow 0$. However, as shown in Fig.~\ref{fig5}(b,c), our $K$-factor estimates are actually higher than in the RC experiments reported in Sec.~\ref{sec_Results}. Therefore, we conclude that the amount of wave energy that is not stirred upon a single encounter with our PM is significant, providing direct evidence for the second identified origin of unstirred wave energy. Moreover, it appears that despite the additional contribution to unstirred wave energy from paths that did not encounter the PM in the RC experiments compared to this AC experiment, this effect is more than outweighed by the fact that many of the paths encountering the PM encounter it multiple times, such that the effect due to the non-zero average of the accessible polarizability values is lower in the RC than in the AC. We corroborate this hypothesis further with numerical results in Sec.~\ref{ss3} below.

\subsection{Numerical Study of Stirring with ``Idealized'' Meta-Atoms}

We now wish to investigate whether the use of ``idealized'' meta-atoms indeed eliminates the second origin of unstirred wave energy, as we hypothesized earlier. Rather than designing a new PM, which is not the goal of this paper focused on PM-stirred RCs, we conveniently explore the impact of the average of the accessible polarizabilities for the meta-atoms in the physics-compliant open-source PhysFad simulator~\cite{faqiri2022physfad}. Whereas finite-element simulations would be prohibitively time-consuming, PhysFad relies on the theory summarized in Sec.~\ref{subsec_gen} which enables the rapid physics-compliant generation of many realizations, a pre-requisite for a thorough systematic investigation of stirring.

For simplicity, as in Ref.~\cite{faqiri2022physfad}, we consider a 2D setup and we work with arbitrary units such that the central operating frequency as well as the medium’s permittivity and permeability are all defined to be unity. These simulations are hence not intended to match a specific experimental setting (although this could in principle be achieved: Ref.~\cite{sol2023experimentally} showed that the PhysFad parameters can be calibrated to a given PM-stirred RC). We begin by qualitatively reproducing the experiment from Sec.~\ref{subsec_AC} with Lorentzian meta-atoms in order to confirm the same qualitative observation of unstirred wave energy although all paths encountered the PM, in order to then see the ``perfect'' stirring if ``idealized'' meta-atoms are used instead. 
In these simulations, we consider a single transmitter and an entire region of interest in which we compute the field. Incidentally, this is an additional benefit of the simulation, since we can easily generate maps of $\hat{K}$ for the region of interest at arbitrarily fine resolution. We have optimized the PhysFad code as detailed in Ref.~\cite{prod2023efficient} and the Appendix to efficiently compute the field realizations underlying the estimation of the $K$-factors for the entire region of interest.
To avoid a strong LOS between the transmitter and our region of interest, an additional line of dipoles acts as LOS block. Whereas in 3D the orientation of the antennas could be controlled to avoid a strong LOS, in 2D there is only one polarization such that this is not possible. Each dipole has a Lorentzian polarizability characterized by the parameters defined in Ref.~\cite{faqiri2022physfad}. Here, we choose $\chi_i=1$ and $\gamma_i^L=0$ for all dipoles, and $f^\mathrm{res}_i$ takes the value of 1 or 5 for dipoles representing meta-atoms in their ``ON'' or ``OFF'' states, respectively, 1 for dipoles representing antennas, and 2 for the dipoles representing environmental scattering objects. In addition, the dipoles representing meta-atoms are backed by a line of static dipoles acting like a metallic reflector at a distance of a quarter wavelength.

In these simulations, we could in principle place the point-like dipoles representing meta-atoms arbitrarily close. However, if the spacing is very small compared to the wavelength and for each realization the configuration of any given dipole representing a meta-atom is arbitrarily chosen, then the overall properties of the PM are almost static because the wave only sees an average property of close-by dipoles which is almost independent of the PM configuration in that case. This insight also generalizes to experiments where the question of the utility of having closely packed deeply sub-wavelength programmable meta-atoms may arise.
Heuristically, we found in our PhysFad simulations that spacing the dipoles representing meta-atoms by 0.3 wavelengths and using them in groups of two (i.e., pairs of neighboring dipoles are always in the same configuration and represent the programmable meta-atoms) optimizes the performance of PM-based stirring.

\begin{figure*}
    \centering
    \includegraphics[width=1\linewidth]{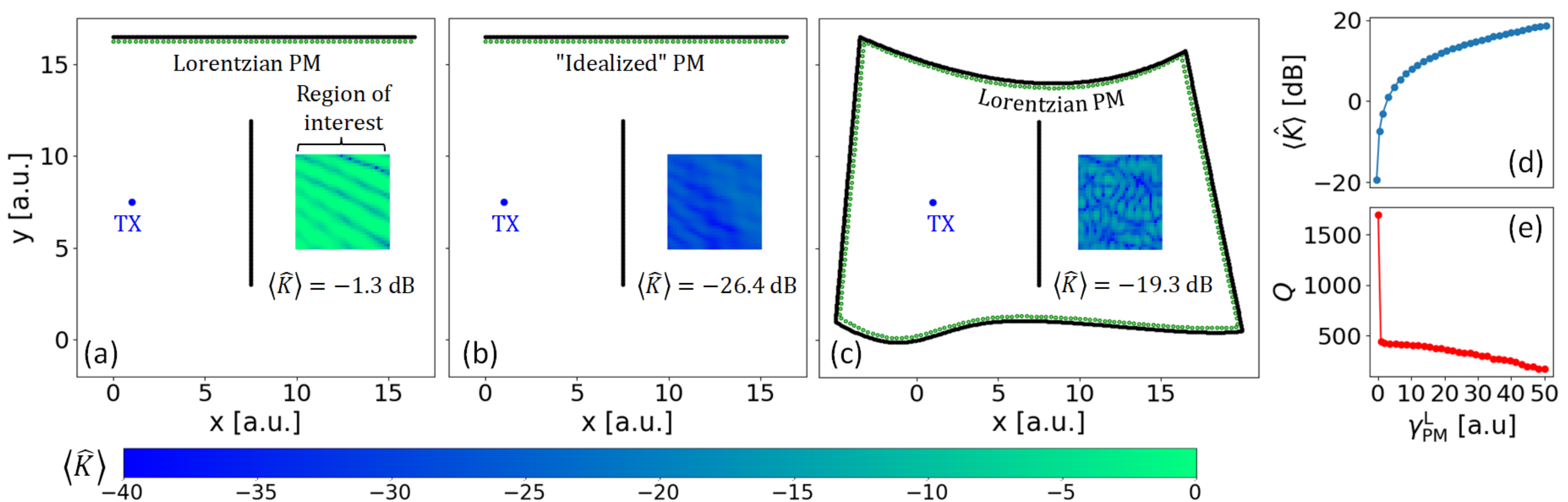}
    \caption{PhysFad simulations investigating the second origin of unstirred wave energy. (a) PhysFad setup qualitatively corresponding to the AC experiment from Fig.~\ref{fig5}(a). Lorentzian meta-atoms are considered and a spatial map of $\hat{K}$ is shown for a region of interest. (b) Same as (a) but with ``idealized'' meta-atoms whose two accessible polarizability values have an average of zero. (c) Fully PM-cladded RC setting with the Lorentzian meta-atoms also used in (a). (d) Dependence of $\langle \hat{K} \rangle$ on the absorption $\gamma_\mathrm{PM}^\mathrm{L}$ of the meta-atoms in the setup from (c). (e) Dependence of the RC's composite $Q$-factor on $\gamma_\mathrm{PM}^\mathrm{L}$ in the setup from (c). }
    \label{fig6}
\end{figure*}

Analogous to the experimental results in the anechoic chamber (Fig.~\ref{fig5}), we observe in the corresponding PhysFad simulation in Fig.~\ref{fig6}(a) that in the case of a PM composed of Lorentzian meta-atoms, a significant amount of unstirred wave energy reaches our region of interest despite having encountered the PM. Averaged over our region of interest, we find $\langle \hat{K} \rangle = -1.4\ \mathrm{dB}$. In addition, we have a rare opportunity to see a spatial map of $\hat{K}$ in Fig.~\ref{fig6}(a). Some structures reminiscent of those seen in field amplitude maps are seen therein, highlighting once again the strong spatial fluctuations of $\hat{K}$.

Then, we replace the Lorentzian meta-atoms by ``idealized'' meta-atoms. Lorentzian meta-atoms have two accessible polarizability values, $\alpha_0$ corresponding to $f^\mathrm{res}_i=5$ and $\alpha_1$ corresponding to $f^\mathrm{res}_i=1$. For the ``idealized'' meta-atoms, we take $\alpha_1$ and $-\alpha_1$ as the two accessible polarizability values, such that there average is exactly zero. Switching from Lorentzian to ``idealized'' meta-atoms essentially removes all unstirred wave energy, as seen in Fig.~\ref{fig6}(b) where $\langle \hat{K} \rangle = -26.4\ \mathrm{dB}$. Thereby, we have provided further clear evidence for the second origin of unstirred wave energy that can be traced back to accessible polarizability values that do not average to zero.

\subsection{Numerical Study of Fully PM-Cladded RC}
\label{ss3}

Next, we asses the impact of reverberation on the second origin of unstirred wave energy. To that end, we consider in Fig.~\ref{fig6}(c) a fully PM-cladded RC (again a scenario that would be challenging to realize experimentally but is easily studied in PhysFad simulations). Unlike our experiments with PM-stirred RCs, in Fig.~\ref{fig6}(c) there are no paths that did not encounter the PM such that any unstirred wave energy can only be of the second origin. Using the same Lorentzian meta-atoms as in Fig.~\ref{fig6}(a), we observe $\langle \hat{K} \rangle = -16.4\ \mathrm{dB}$ in Fig.~\ref{fig6}(c). In other words, despite the non-zero average of the accessible polarizabilities, there is essentially no unstirred wave energy in this fully PM-cladded RC. The key difference between the AC setting from Fig.~\ref{fig6}(a) and the RC setting from Fig.~\ref{fig6}(c) is that in the former case the waves encounter the PM only once whereas they encounter it countless times in the latter case. Even though only a portion of the wave energy is stirred upon a single encounter with the PM, no significant amount of energy can avoid to be eventually stirred after multiple bounces. The longer the reverberation time, the more sensitive the wave field becomes to any perturbation and the better the PM can stir it (this ``generalized interferometric sensitivity'' also enables sub-wavelength sensing~\cite{del2021deeply}). 
This finding is in line with our experimental observations where PM-based stirring performed better in the RC than the AC (despite the fact that in our RC experiments the RC was not fully PM-cladded such that there was additionally unstirred wave energy due to effect (i): paths that never encounter the PM). Consequently, minimizing attenuation in the RC is of significant importance to facilitate PM-based stirring.

Again, we get a rare insight into the spatial map of $\hat{K}$ in Fig.~\ref{fig6}(c), and in this RC setting it features speckle patterns, again reminiscent of field amplitude maps inside chaotic RCs. Once more, the strong spatial fluctuations of $\hat{K}$ are hence apparent.

\subsection{Numerical Study of Effect of PM Absorption}

Finally, to further evidence the important role of the reverberation time for PM-based RC stirring, we study in Fig.~\ref{fig6}(d) how the value of $\langle \hat{K} \rangle$ evolves as we increase the level of absorption of the meta-atoms. As seen in Fig.~\ref{fig6}(e), more absorption by the meta-atoms decreases the RC's composite quality factor, and, as expected, is seen to result in more wave energy remaining unstirred.

\section{Conclusion}
\label{sec_Conclusion}

To summarize, we have experimentally, numerically and theoretically explored the potential of PM-stirred RCs for OTA and EMC tests. Specifically, our goal was to emulate Rician fading with electronically adjustable $K$-factor for MIMO systems, which is not easily achieved with existing RC-based techniques. To the best of our knowledge, our results are the first large-scale investigation of PM-stirred RCs and the first time that physical fields rather than only stirred field components are analyzed. Our results confirm the feasibility of implementing adjustable Rician fading for MIMO systems within a range of $K$-factor values relevant to many indoor settings. At the same time, our results highlight the strong spatial fluctuations of the $K$-factor and reveal strong constraints on the range of accessible spatially averaged $K$-factor values. In particular, we thoroughly investigated the origins of unstirred wave energy that impose a lower bound on the $K$-factor achievable in a given setting. Besides the obvious origin related to paths that never encounter the PM, we identified a second origin that has to date not been reported in the literature to the best of our knowledge. This second effect is concerned with wave energy that is only partially stirred despite travelling along the paths that do encounter the PM. We traced this possibly surprising effect back to the non-zero average of the accessible polarizability values of the programmable meta-atoms. The effect can also be interpreted in terms of the meta-atoms' non-zero ``structural'' scattering cross-section. 
We established that longer reverberation times, increasing the number of encounters with the PM, effectively alleviate the contributions of both origins of unstirred wave energy. Looking forward, our work highlights the importance of minimizing attenuation within the RC and seeking ``idealized'' programmable meta-atoms (i.e., zero average of accessible polarizability values) for
future efforts aimed at efficiently suppressing unstirred wave energy to emulate Rayleigh-like conditions in PM-stirred RCs for OTA or EMC tests.

\section*{Acknowledgments}
The authors gratefully acknowledge the assistance of F.~Boutet, C.~Guitton, L.~Le~Coq, J.~Lorandel and J. Sol with various aspects of the experimental work.

\appendix[Algorithmic Details for PM-Stirring in PhysFad]
\label{appendix_physfad}

In order to estimate the $K$-factor for each grid point within a region of interest, the fields at these grid points corresponding to $m$ random PM configurations must be evaluated. 

Following the notation and assumptions from Ref.~\cite{faqiri2022physfad}, the field $E_\mathrm{loc}(\mathring{\mathbf{r}})$ at a given location $\mathring{\mathbf{r}}$ (assumed not to coincide with any of the dipole locations $\mathbf{r}_i$) is given by
\begin{equation}
    E_\mathrm{loc}(\mathring{\mathbf{r}}) = \sum_{i=1}^N G(\mathring{\mathbf{r}},\mathbf{r}_i)p_i,
    \label{eq:dipole_fields}
\end{equation}
where $N$ is the number of dipoles, $p_i$ is the dipole moment of the $i$th dipole, and $G(\mathring{\mathbf{r}},\mathbf{r}_i)$ is the Green's function between the locations $\mathbf{r}_i$ and $\mathring{\mathbf{r}}$.

The dipole moments are evaluated via
\begin{equation}
    \mathbf{p} = \mathbf{W}^{-1} \mathbf{E}^\mathrm{ext},
    \label{eq11}
\end{equation}
where $\mathbf{p}=[p_1,\dots,p_N]$ and $\mathbf{E}^\mathrm{ext}=[E^\mathrm{ext}_1,\dots,E^\mathrm{ext}_N]$, 
with $E^\mathrm{ext}_i$ being the external field incident on the location of the $i$th dipole such that $E^\mathrm{ext}_i = 0 \ \forall \ i \in \mathcal{R} \cup \mathcal{S} \cup \mathcal{E}$.

Because $\mathbf{W}$ only differs regarding parts of its diagonal for different PM configurations, the Woodbury identity can be applied to update a previously evaluated $\mathbf{W}^{-1}$ instead of performing an entire new matrix inversion, as detailed in Sec.~IV.A of Ref.~\cite{prod2023efficient}. In the following, the tilde denotes updated quantities. 
Since all entries of $\mathbf{E}^\mathrm{ext}$ are zero except for those corresponding to indices contained in the set $\mathcal{T}$, it is in fact sufficient to evaluate the update of the $\mathcal{NT}$ block of  $\mathbf{W}^{-1}$, where $\mathcal{N} = \mathcal{T} \cup \mathcal{R} \cup \mathcal{E} \cup \mathcal{S}  $. Hence, following Ref.~\cite{prod2023efficient}, the computationally most efficient approach consists in solely evaluating 
\begin{multline}
\left[\tilde{\mathbf{W}}^{-1}\right]_{\mathcal{NT}} = \left[\left(\mathbf{W}+\mathbf{{U}C{V}}\right)^{-1}\right]_{\mathcal{NT}} = \left[\mathbf{W}^{-1}\right]_{\mathcal{NT}} - \\ \left[\mathbf{W}^{-1}\right]_{\mathcal{NM}}\left(\mathbf{C}^{-1}\!+\!\left[\mathbf{W}^{-1}\right]_\mathcal{MM}\right)^{-1}\left[\mathbf{W}^{-1}\right]_{\mathcal{MT}},
\label{eq:woodbury}
\end{multline}
where $\mathcal{M}$ denotes the set of dipole indices whose configuration changed and $\tilde{\mathbf{W}}$ denotes the updated interaction matrix. As detailed in Ref.~\cite{prod2023efficient}, the size of the set $\mathcal{M}$ can be limited to at most $\lfloor N_\mathrm{S}/2 \rfloor$ if two complementary reference PM configurations are chosen.

To evaluate the field $ \tilde{E}_\mathrm{loc}(\mathring{\mathbf{r}},\mathbf{\tilde{c}})$ for a new PM configuration $\mathbf{\tilde{c}}$ given a previously evaluated $\mathbf{W}^{-1}$ for a PM configuration $\mathbf{c}$, we proceed as follows. First, we evaluate the updated $\mathcal{NT}$ block of the corresponding inverse interaction matrix denoted by $\left[\tilde{\mathbf{W}}^{-1}\right]_{\mathcal{NT}}$ with Eq.~(\ref{eq:woodbury}). Second, we evaluate the updated dipole moments $\mathbf{\tilde{p}} = \left[\tilde{\mathbf{W}}^{-1}\right]_{\mathcal{NT}} \mathbf{E}^\mathrm{ext}_\mathcal{T} $ following Eq.~(\ref{eq11}). Since we only consider a single transmitter, we work with $\mathbf{E}^\mathrm{ext}_\mathcal{T}=1$ for simplicity and without loss of generality.  
Third, we evaluate $\tilde{E}_\mathrm{loc}(\mathring{\mathbf{r}}) = \sum_{i=1}^N G(\mathring{\mathbf{r}},\mathbf{r}_i)\tilde{p}_i$, where $\tilde{p}_i$ is the $i$th entry of $\mathbf{\tilde{p}}$. Based on the values of $ \tilde{E}_\mathrm{loc}(\mathring{\mathbf{r}},\mathbf{\tilde{c}})$ for $m$ random realizations of $\mathbf{\tilde{c}}$, we can then evaluate the $K$-factor estimate $\hat{K}$ via Eq.~(\ref{eq_philippe}).

The computational efficiency of the $K$-factor estimation can be optimized by representing it as a tensor product with two outer dimensions (positions $\mathring{\mathbf{r}}$ and PM configurations $\mathbf{c}$), reduced through summation along the third inner dimension of the dipoles indices $i$.

\bibliographystyle{IEEEtran}

\end{document}